# High Order Control Lyapunov Function - Control Barrier Function - Quadratic Programming Based Autonomous Driving Controller for Bicyclist Safety

Haochong Chen, Xincheng Cao, Levent Guvenc, Bilin Aksun-Guvenc

Automated Driving Lab, Ohio State University

## Abstract

Ensuring the safety of Vulnerable Road Users (VRUs) is a critical challenge in the development of advanced autonomous driving systems in smart cities. Among vulnerable road users, bicyclists present unique characteristics that make their safety both critical and also manageable. Vehicles often travel at significantly higher relative speeds when interacting with bicyclists as compared to their interactions with pedestrians which makes collision avoidance system design for bicyclist safety more challenging. Yet, bicyclist movements are generally more predictable and governed by clear traffic rules as compared to the sudden and sometimes erratic pedestrian motion, offering opportunities for model-based control strategies. To address bicyclist safety in complex traffic environments, this study proposes and develops a High-Order Control Lyapunov Function–High-Order Control Barrier Function–Quadratic Programming (HOCLF-HOCBF-QP) control framework. Through this framework, CLFs constraints guarantee system stability so that the vehicle can track its reference trajectory, whereas CBFs constraints ensure system safety by letting vehicle avoiding potential collisions region with surrounding obstacles. Then by solving a QP problem, an optimal control command that simultaneously satisfies stability and safety requirements can be calculated. Three key bicyclist crash scenarios recorded in the Fatality Analysis Reporting System (FARS) are recreated and used to comprehensively evaluate the proposed autonomous driving bicyclist safety control strategy in a simulation study. Simulation results demonstrate that the HOCLF-HOCBF-QP controller can help the vehicle perform robust, and collision-free maneuvers, highlighting its potential for improving bicyclist safety in complex traffic environments.

## Introduction

Ensuring the safety of Vulnerable Road Users (VRU) has already become a become a major concern for modern cities. Autonomous driving vehicles with VRU safety algorithms for collision avoidance present a possible solution [1], [2], [3]. Compared with vehicles, pedestrians and cyclists are far more fragile and therefore suffer disproportionately high rates of injury and fatality in traffic accidents. Among all types of VRUs, bicyclists face particularly high risks, suffering disproportionately high fatality rates in collisions, which may need extra attention to improve their safety and reduce accident severity.

While pedestrian movements are often difficult to predict, they often stay on sidewalks or designated crosswalks, which limits their direct interaction with vehicles [4], [5], [6]. Bicyclists, in contrast, typically travel at higher speeds and need to frequently share roadways with other road users. Compared to pedestrians, bicyclist movements are generally more predictable and governed by clear traffic rules which offers opportunities for modeling their behavior pattern. However, their small size and limited visibility increases the likelihood of being overlooked, especially in blind spots. Moreover, the need to occupy full lanes or bike-specific lanes let bicyclists have more frequent and direct interactions with vehicles, which increases their risk of collision with other road users. This highlights the need for advanced autonomous driving systems that explicitly address bicycle safety.

Autonomous Driving Systems (ADS) can address aforementioned challenges and enhance road safety, especially VRU safety, by reducing human-error-related accidents [7]. Extensive research has already been conducted in the autonomous driving field to help vehicles navigate safely and efficiently [8-10]. Current approaches can be broadly categorized into two types. The first is end-to-end, model-free learning, where an autonomous driving agent directly maps sensory inputs to control actions. Bojarski et al. were among the first to propose a convolutional neural network based end-to-end framework for autonomous driving systems (ADS) [11]. Kendall et al. then applied deep reinforcement learning (DRL) methods to training an autonomous driving agent [12]. Building on this, subsequent research has expanded end-to-end applications in ADS with several innovations [13-15], including uncertainty-weighted decision transformer architecture for complex roundabout scenarios [16] and hierarchical reinforcement learning (H-REIL) to better balance safety and efficiency in near-accident scenarios [17]. Compared to traditional modular based approaches, end-to-end method allows for optimization across the entire processing pipeline and offers the potential for improved overall performance. Nevertheless, the lack of explicitly defined safety constraints can be problematic, as such systems might perform unsafe behaviors in rare or safety-critical scenarios.

The second category is modular, model-based design, which remains the mainstream in autonomous driving systems development [18-22]. In this paradigm, trajectory planning and collision avoidance are treated as dedicated modules, often formulated as optimization problems or addressed using machine learning based method to generate safe and feasible paths. Ames et al. formulated path following and collision avoidance as constraints using CLF-CBF-QP to ensure both efficient and safe navigation [23-26]. Nageshrao et al. proposed to apply short-horizon safety mechanisms for highway autonomous driving [27]. Zhang et al. introduced a bootstrapping based demonstration policy to accelerate training of high-level autonomous driving decision-making agent [28]. This modular design allows researchers to isolate and improve individual components, ultimately achieving better overall system reliability. In such frameworks, the low-level controller is typically a separate module that receives commands from the high-level decision-making layer and executes precise control actions to ensure safe and efficient navigation. A well-designed low-level controller plays an important role in autonomous driving systems. It ensures that the vehicle can accurately execute high-level decisions leading to precise path tracking under various traffic conditions.

Traditional low-level controllers, such as pure-pursuit controller [29], [30], Stanley controller [31], disturbance observer (DOB) [32], parameter-space multi-objective PID control [33], model predictive control (MPC) [34], [35], [36] and other optimization based controller [37], [38], [39], [40], [41], primarily focus on accurate path tracking. While effective under nominal conditions, they often lack the flexibility to handle emergencies or compensate for errors when the high-level agent makes unsafe decisions. In such cases, the inability to adapt may directly lead to traffic accidents. Therefore, it is essential to enhance traditional path-tracking frameworks with an additional safety layer that can actively enforce collision avoidance and protect VRUs safety. This paper introduces a HOCLF-HOCBF-QP-based low-level controller to handle path planning and collision avoidance in autonomous driving. The key contributions are outlined below.

- A novel HOCLF–HOCBF-based low-level controller is developed for vehicle dynamic models, providing accurate path tracking in normal collision free condition and reliable collision avoidance around VRUs.

- This low-level controller takes advantage of the optimization-based approach. The HOCLF constraint in the optimization ensures stability and accurate path tracking of the autonomous vehicle, while the HOCBF constraint guarantees safety by preventing potential collisions with other vehicles and obstacles.

- Real FARS bicyclist crash cases are recreated for simulation studies which can evaluate overall collision avoidance performance of the proposed CLF-CBF based control framework.

The remainder of this paper is structured as follows. The methodologies employed in this study is described in section 2, which includes the single-track lateral vehicle dynamic model and the formulation of the low-level HOCLF–HOCBF–QP controller. Section 3 presents the simulation results obtained from the proposed framework, including trajectory-tracking test using the HOCLF–QP controller, collision avoidance test of the low-level HOCLF–HOCBF–QP controller, and overall evaluation based on real-world FARS bicyclist crash scenarios. A detailed discussion of these results is also provided in this section. Finally, Section 4 concludes the paper and discusses potential future research.

## Methodology

### Single-Track Lateral Vehicle Dynamic Model

In this paper, a linear single-track lateral vehicle dynamic model is utilized for controller design and vehicle simulation. As shown in Figure 1, the model describes planar lateral motion of the vehicle at constant speed, with wheel side slip angles included for realism. The complete derivation of this vehicle model can be found in Chapter 2 of [42]. Equations 1 define the state-space equations of the vehicle dynamics where the state variables include the vehicle side slip angle ($\beta$) and yaw rate ($r$), while the inputs include the front and rear steering angles ($\delta_f$ and $\delta_r$). A yaw moment disturbance ($M_{zd}$) is also incorporated as an external disturbance. Equations 2 and 3 are used to calculate the vehicle's position based on vehicle speed and the yaw angle ($\psi$). Table 1 summarizes the model parameters, which are adopted from [43].

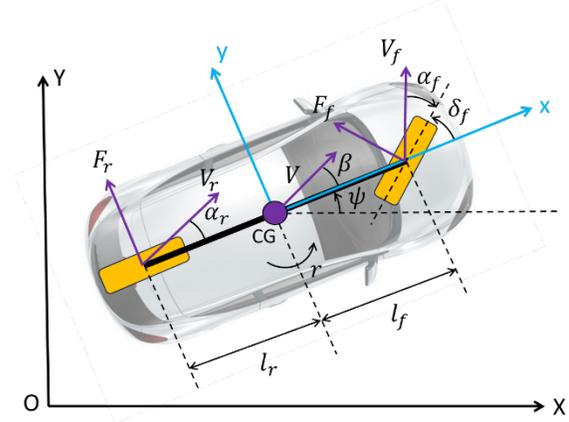

Figure 1. Linear single-track lateral vehicle dynamic model.

Table 1. Lateral model parameters [43].

| Symbol | Parameter |
|---|---|
| $X, Y$ | Earth-fixed frame coordinate |
| $x, y$ | Vehicle-fixed frame coordinate |
| $V$ | Center-of-gravity (CG) velocity |
| m | Mass |
| $I_z$ | Yaw moment of inertia |
| $\beta$ | Side-slip angle |
| $\psi$ | Yaw angle |
| r | Yaw rate |
| $M_{zd}$ | Yaw disturbance moment |
| $\delta_f, \delta_r$ | Front & rear wheel steer angle |
| $\alpha_f, \alpha_r$ | Front & rear tire slip angle |
| $C_f, C_r$ | Front & rear tire cornering stiffness |
| $l_f, l_r$ | Distance between CG and front & rear axle |
| $V_f, V_r$ | Front & rear axle velocity |
| $F_f, F_r$ | Front & rear lateral tire force |

$$\begin{bmatrix} \dot{\beta} \\ \dot{r} \end{bmatrix} = \begin{bmatrix} \frac{-C_f - C_r}{mV} & -1 + \frac{C_r l_r - C_f l_f}{mV^2} \\ \frac{C_r l_r - C_f l_f}{I_z} & \frac{-C_f l_f^2 - C_r l_r^2}{I_z V} \end{bmatrix} \begin{bmatrix} \beta \\ r \end{bmatrix} + \begin{bmatrix} \frac{C_f}{mV} & \frac{C_r}{mV} \\ \frac{C_f l_f}{I_z} & \frac{C_r l_r}{I_z} \end{bmatrix} \begin{bmatrix} \delta_f \\ \delta_r \end{bmatrix} + \begin{bmatrix} 0 \\ \frac{1}{I_z} \end{bmatrix} M_{zd} \quad (1)$$

$$\Delta x = \int_0^{t_f} v\cos(\beta + \psi)\, dt \quad (2)$$

$$\Delta y = \int_0^{t_f} v\sin(\beta + \psi)\, dt \quad (3)$$

In this model, several assumptions are made including front-steering vehicle ($\delta_r = 0$) and zero yaw moment disturbances ($M_{zd} = 0$). After making these assumptions, the system can be represented by a five-degree-of-freedom (5-DOF) lateral vehicle dynamic model. The state-space equation for the proposed 5-DOF vehicle lateral dynamic model is given by

$$\begin{bmatrix} \dot{\beta} \\ \dot{r} \\ \dot{x} \\ \dot{y} \\ \dot{\psi} \end{bmatrix} = \begin{bmatrix} A_{11} * \beta + A_{12} * r \\ A_{21} * \beta + A_{22} * r \\ v * \cos(\beta + \psi) \\ v * \sin(\beta + \psi) \\ r \end{bmatrix} + \begin{bmatrix} B_1 \\ B_2 \\ 0 \\ 0 \\ 0 \end{bmatrix} \delta_f \quad (4)$$

where $A_{11} = \frac{-C_f - C_r}{mv}$, $A_{12} = -1 + \left(\frac{C_r l_r - C_f l_f}{mv^2}\right)$, $A_{21} = \frac{C_r l_r - C_f l_f}{I_z}$, $A_{22} = \frac{-C_f l_f^2 - C_r l_r^2}{I_z V}$, $B_1 = \frac{C_f}{mv}$ and $B_2 = \frac{C_f l_f}{I_z}$. Unlike simplified lateral dynamic models demonstrated in Equation (1), which consider only internal vehicle states like sideslip angle and yaw rate. The proposed model incorporates the vehicle's position and heading ($x, y, \psi$). This augmentation provides a more intuitive geometric understanding of the vehicle's motion, which is especially useful for developing HOCLF–HOCBF–based controllers, where reference tracking and obstacle avoidance are formulated in the global coordinate frame.

## *Control Lyapunov Functions and Control Barrier Functions*

### Preliminary

In this section, we first introduce CLF and CBF. For a control-affine system, the dynamics can be written as:

$$\dot{x} = F(x, u) = f(x) + g(x)u \quad (5)$$

where $x \in \mathbb{R}^n$ is system state, $u \in \mathbb{R}^m$ is control input, $f(x)$, $g(x)$ are locally Lipschitz functions. $f(x)$ represents the system dynamics, while $g(x)$ describes how control inputs affect the system.

The control goal is to let the vehicle navigate to the desired destination, which requires stabilizing the system around an equilibrium point. A common approach is to use a CLF as a constraint in an optimization-based controller, ensuring Lyapunov stability. Specifically, CLF is a continuously differentiable function $V(x): R^n \to R$, that satisfies certain inequalities with positive constants $c_1 > 0$, $c_2 > 0$, $c_3 > 0$ such that for $\forall x \in \mathbb{R}^n$,

$$c_1 \|x\|^2 \leq V(x) \leq c_2 \|x\|^2 \quad (6)$$

and also,

$$\inf_{u \in U} [L_f V(x) + L_g V(x) u + c_3 V(x)] \leq 0 \quad (7)$$

where $L_f$ and $L_g$ represents Lie derivatives along $f(x)$ and $g(x)$. If these conditions hold, then $V(x)$ is a CLF for the system ensuring global and exponential stabilization.

Besides stability, another key control objective is collision avoidance when obstacles are present. To ensure safety, CBFs are used. A CBF defines a safe set of states and enforces the system to keep staying inside this set over time. Like CLFs, CBFs can be added as constraints in an optimization-based framework, such as quadratic programming (QP). Formally, a CBF is defined by a continuously differentiable function $h(x): R^n \to R$, and the corresponding specifies the safe set as:

$$C = \{x \in R^n: h(x) \geq 0\} \quad (8)$$

The function $h(x)$ is considered a CBF if there exists a control input $u \in R^n$ such that the following condition holds for $\forall x \in C$:

$$\sup_{u \in U} [L_f h(x) + L_g h(x) u + \alpha(h(x))] \geq 0 \quad (9)$$

where $\alpha$ is a class-$\kappa$ function. The detailed derivation and proof of CLF and CBF can be found in [39].

### HOCLF and HOCBF Definition

In more complicated systems, where safety constraints involve higher-order derivatives of the system states, traditional first order CLFs and CBFs are no longer sufficient. To address this, High-Order CLFs (HOCLFs) and High-Order CBFs (HOCBFs) are introduced. The relative degree of a HOCLF or HOCBF is defined as the number of times the function must be differentiated along the system dynamics before the control input $u$ explicitly appears.

Specifically, let HOCLF be a $d$ th-order continuously differentiable function $V(x): R^n \to R$. Then, define $\phi_0(x) = V(x)$ and construct a sequence of functions $\phi_i(x): R^n \to R, i \in \{1, \dots, d\}$ such that:

$$\phi_i(x) = \dot{\phi}_{i-1}(x) + \alpha_i(\phi_{i-1}(x)), i \in \{1, \dots, d\} \quad (10)$$

where each $\alpha_i$ is a class-$\kappa$ function. If there exist $\alpha_d$ such that $\forall x \neq 0_n$

$$\inf_{u \in U} [L_f^d V(x) + L_g L_f^{d-1} V(x) u + S(h(x)) + \alpha_d(\phi_{d-1}(x))] \leq 0 \quad (11)$$

where $L_f$ and $L_g$ represent Lie derivatives along $f(x)$ and $g(x)$, then $V(x)$ is a HOCLF for the system, guaranteeing global and exponential stabilization.

Similarly, HOCBF is an $r$ th-order continuously differentiable function $h(x): R^n \to R$. Then, define $\Psi_0(x) = h(x)$ and a sequence of functions $\Psi_i(x): R^n \to R, i \in \{1, \dots, r\}$:

$$\Psi_i(x) = \dot{\Psi}_{i-1}(x) + \beta_i(\Psi_{i-1}(x)), i \in \{1, \dots, r\} \quad (12)$$

where each $\beta_i$ is a class-$\kappa$ function. Correspondingly, define a sequence of sets $C_i, i \in \{1, \dots, r\}$:

$$C_i(x) = \{x \in \mathbb{R}^n: \Psi_{i-1}(x) \geq 0\}, i \in \{1, \dots, r\} \quad (13)$$

If there exists $\beta_r$ and a control input $u \in R^n$ such that the following condition holds for $\forall x \in C_1 \cap C_2 \cap \dots \cap C_i$

$$\sup_{u \in U} [L_f^m h(x) + L_g L_f^{m-1} h(x) u + S(h(x)) + \alpha_i(\Psi_{i-1}(x))] \geq 0 \quad (14)$$

where $L_f$ and $L_g$ denote Lie derivatives along $f(x)$ and $g(x)$, then $h(x)$ is a HOCBF for the system which can guarantee safety. The detailed derivation and proof of HOCLF and HOCBF can be found in [40], [44].

### HOCLF and HOCBF Design

The primary objective for the low-level path tracking controller design is to compute safe and effective control inputs that enable the vehicle to follow a desired path from the start to the destination while avoiding collisions. As the first step, we develop the path-following component based on the aforementioned HOCLF framework.

$$V(x) = (x - x_g)^2 + (y - y_g)^2 \quad (15)$$

$$\dot{V}(x,u) = \mathcal{L}_f V(x) + \mathcal{L}_g V(x)u = 2v\cos(\beta + \psi)(x - x_g) + 2v\sin(\beta + \psi)(y - y_g) \tag{16}$$

$$\mathcal{L}_f^2 V(x) = \nabla\left(\mathcal{L}_f V(x)\right)^T f(x) = -2v\sin(\beta + \psi)(x - x_g)(A_{11}\beta + A_{12}r) + 2v\cos(\beta + \psi)(y - y_g)(A_{11}\beta + A_{12}r) + 2v^2 - 2v\sin(\beta + \psi)(x - x_g)r + 2v\cos(\beta + \psi)(y - y_g) \tag{17}$$

$$\mathcal{L}_g\mathcal{L}_f V(x)u = \nabla\left(\mathcal{L}_f V(x)\right)^T g(x) = -2v\sin(\beta + \psi)(x - x_g)B_1\delta_f + 2v\cos(\beta + \psi)(y - y_g)B_1\delta_f \tag{18}$$

Equations 15–18 describe the HOCLF formulation and its derivatives, where $[x,y]$ denote the vehicle's current coordinates and $[x_g, y_g]$ represent the destination. The coefficients $A_{11}, A_{12}, A_{21}, A_{22}, B_1, B_2$ are derived from the previously defined vehicle dynamics model. HOCLF is used to ensure system stability, which can force the vehicle navigating toward target waypoint on the desired path. The concept is straightforward: the vehicle's position should gradually converge to the goal coordinates. To achieve this, we construct a Lyapunov candidate function $V(x)$ that measures the squared distance between the current position and the goal. Equation (19), then, shows the HOCLF constraint.

$$\mathcal{L}_f^2 V(x) + \mathcal{L}_g\mathcal{L}_f V(x)u + \alpha_1\left(\dot{V}(x,u)\right) + \alpha_2(V(x)) \leq \delta \tag{19}$$

Here, $\alpha_1$ and $\alpha_2$ are class-$\kappa$ functions and $\delta$ is a slack variable, which are implemented as positive constant gains. This constraint ensures that the control input $u$ consistently drives the system towards the goal in a stable and controlled manner.

Similarly, we design the collision-avoidance component using HOCBFs. Define the barrier candidate function:

$$h(x) = (x - x_o)^2 + (y - y_o)^2 - r_o^2 \tag{20}$$

$$\dot{h}(x,u) = \mathcal{L}_f h(x) + \mathcal{L}_g h(x)u = \mathcal{L}_f h(x) = 2v\cos(\beta + \psi)(x - x_o) + 2v\sin(\beta + \psi)(y - y_o) \tag{21}$$

$$\mathcal{L}_f^2 h(x) = \nabla\left(\mathcal{L}_f h(x)\right)^T f(x) = -2v\sin(\beta + \psi)(x - x_o)(A_{11}\beta + A_{12}r) + 2v\cos(\beta + \psi)(y - y_o)(A_{11}\beta + A_{12}r) + 2v^2 - 2v\sin(\beta + \psi)(x - x_o)r + 2v\cos(\beta + \psi)(y - y_o)r \tag{22}$$

$$\mathcal{L}_g\mathcal{L}_f h(x)u = \nabla\left(\mathcal{L}_f h(x)\right)^T g(x)u = -2v\sin(\beta + \psi)(x - x_o)B_1\delta_f + 2v\cos(\beta + \psi)(y - y_o)B_1\delta_f \tag{23}$$

Equations 20–23 describe the HOCBF formulation and its derivatives, where $[x,y]$ denote the vehicle's current coordinates and $[x_o, y_o]$ represent the coordinates of the obstacles. The coefficients $A_{11}, A_{12}, A_{21}, A_{22}, B_1, B_2$ were defined in the previously defined vehicle dynamics model. HOCBF is used to ensure system safety, in order to keep the vehicle away from potential collision with nearby obstacles. The concept is straightforward: the vehicle's position should not enter a circular danger zone of radius $r_o$ centered at $[x_o, y_o]$. To achieve this, we construct a barrier candidate function $h(x)$ which quantifies whether the center of the vehicle is entering the dangerous zone or not. For each obstacle, a corresponding HOCBF is needed. The HOCBF constraint is formulated as:

$$\mathcal{L}_f^2 h(x) + \mathcal{L}_g\mathcal{L}_f h(x)u + \alpha_3\left(\mathcal{L}_f h(x)\right) + \alpha_4(h(x)) \geq 0 \tag{24}$$

Here, $\alpha_1$ and $\alpha_2$ are class-$\kappa$ functions, which are implemented as positive constant gains. This constraint ensures that the control input $u$ can keep the system away from the unsafe region with obstacles.

### HOCLF-HOCBF-QP Formulation

To integrate the previously designed HOCLF and HOCBF into a unified control framework, we formulate a QP that incorporates both HOCLF and HOCBF as inequality constraints. This optimization-based approach allows the controller to compute control inputs that not only guide the system towards the desired destination but also ensure safety by preventing vehicle entry into unsafe regions around obstacles.

The QP is defined as follows:

$$u^* = \arg\min_{u,\delta} \|u - u_{ref}\|^2 + q\delta^2 \tag{25}$$

$$s.t \quad \mathcal{L}_f^2 V(x) + \mathcal{L}_g\mathcal{L}_f V(x)u + \alpha_1\left(\dot{V}(x,u)\right) + \alpha_2(V(x)) \leq \delta$$

and

$$\mathcal{L}_f^2 h(x) + \mathcal{L}_g\mathcal{L}_f h(x)u + \alpha_3\left(\mathcal{L}_f h(x)\right) + \alpha_4(h(x)) \geq 0, \text{for each obstacle}$$

Here, $u$ is the control input and $u_{ref}$ is a nominal reference input, which is usually set to zero to minimize control effort. The slack variable $\delta$ provides flexibility by allowing temporary relaxation of the HOCLF constraint in cases of conflict with the HOCBF. The penalty weight $q > 0$ determines the trade-off between control performance and constraint violation.

## Experiments and Results
### HOCLF-HOCBF Based Optimization Controller

**Path Tracking Performance Test**

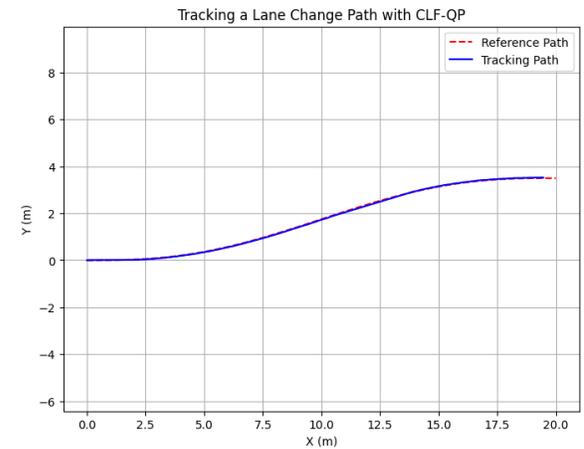

Figure 3. Path tracking test simulation results.

Figure 3 illustrates the reference trajectory-tracking results of the proposed HOCLF–QP controller. The red dashed curve denotes the desired single-lane-change path, whereas the blue solid curve is the actual motion of the ego-vehicle controlled by proposed HOCLF–QP controller. As shown in the figure, the vehicle tracks the reference trajectory with high precision and minimal lateral tracking error across the entire path. In this design, the HOCLF serves as a stability constraint, which allows ego-vehicle to accurately follow the desired path. This result shows that the proposed HOCLF-QP controller has reliable low-level path tracking capability, making it a suitable low-level controller in the autonomous driving framework.

**Collision Avoidance Performance Test with Static Obstacle**

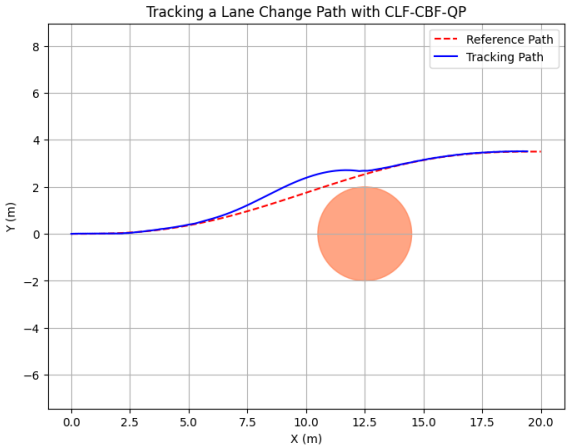

Figure 4. Reference trajectory tracking test simulation results

Figure 4 illustrates the trajectory tracking performance of the HOCLF-HOCBF-QP controller in traffic environment with a static obstacle. The red dashed line shows the predefined reference path for the single-lane-change maneuver, while the orange circle marks a static obstacle intentionally placed close to this path. The blue solid line shows the actual trajectory of the ego-vehicle under the HOCLF-HOCBF-QP controller.

In contrast to the HOCLF-QP control discussed in the previous section, which only performs path tracking, the inclusion of the HOCBF enables the vehicle to modify its trajectory to avoid a potential collision while still maintaining adherence to the reference path after bypassing the obstacle. The HOCBF serves as a safety constraint, ensuring that the system state remains within a safe set even when the original reference path would have resulted in a collision. The temporary deviation from the reference trajectory observed near the obstacle reflects the collision avoidance capability introduced by the HOCBF. Once the obstacle is cleared, the vehicle smoothly converges back to the reference path, demonstrating the controller's effectiveness in balancing safety with path tracking stability.

Figure 5 presents the reference-point tracking simulation results of proposed HOCLF–HOCBF–QP controller in an environment containing multiple static obstacles. The ego vehicle begins from an arbitrary initial position and aims to reach the target destination, which is represented by green dots, while avoiding collisions with circular obstacles, which is marked by orange color. The dashed curve demonstrates the actual motion of the ego-vehicle under the control of the HOCLF–HOCBF–QP controller. In contrast to the previous case, which primarily evaluated the controller's ability to continuously track a sequence of waypoints (similar to pure pursuit), this scenario focuses on testing the controller's capability to track a single reference point while ensuring safe obstacle avoidance and successful arrival at the destination.

It is observed from the figure that the vehicle can successfully reach the goal while smoothly passing around all three obstacles. This behavior demonstrates that the HOCBF constraints effectively enforce safety by keeping the system state away from unsafe regions, while the HOCLF ensures stable tracking of the reference target. The results highlight the controller's capability to achieve both collision avoidance and accurate goal tracking in complex driving environments.

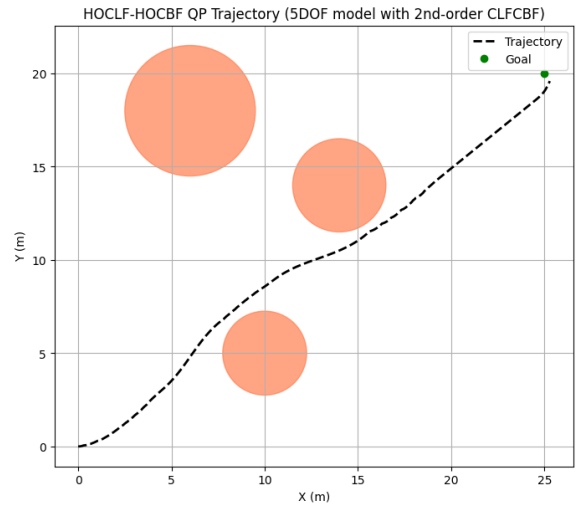

Figure 5. Reference point tracking test simulation results

Overall, these two cases confirm the effectiveness of the HOCLF-HOCBF-QP controller in static obstacle traffic scenarios, showing robust performance in both reference path tracking and reference point tracking, and highlighting its suitability as a low-level module within an autonomous driving system.

**Collision Avoidance Performance Test with Dynamic Obstacle**

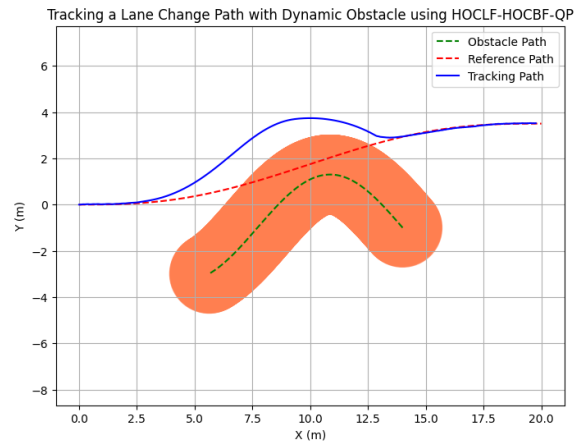

Figure 6. Reference trajectory tracking test simulation results with dynamic obstacles.

To further test the collision-avoidance performance of proposed low-level HOCLF–HOCBF–QP controller, a traffic environment containing a dynamic obstacle that follows a predefined path is constructed for testing. As shown in Figure 6, the ego-vehicle can follow its reference path both before and after interacting with the obstacle, while executing a collision avoidance maneuver when encountering the obstacle. The vehicle temporarily deviates from the reference trajectory to maintain safety and then smoothly converges back after bypassing the obstacle. This demonstrates the controller's ability to ensure safe navigation in the presence of moving obstacles while preserving accurate path tracking.

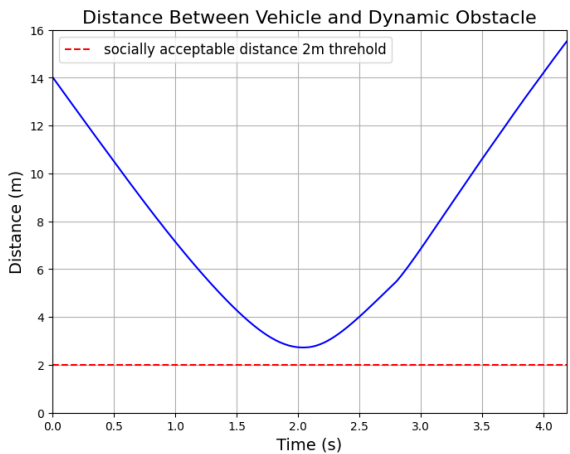

Figure 7. Distance between ego-vehicle and dynamic obstacles

Figure 7 presents the time plot of the distance between the ego-vehicle and the dynamic obstacle. The minimum separation occurs at approximately 2.0 seconds, corresponding to the point of closest interaction between ego-vehicle and obstacle. Importantly, this minimum distance remains above the predefined 2-meter safety threshold, which is generally regarded as a socially acceptable lower bound for safe operation in typical low-speed, urban driving conditions [45]. These results confirm that the HOCLF-HOCBF-QP controller not only avoids collisions with static obstacles but also can maintain safe distance when encountering dynamic obstacles.

The simulation results presented before demonstrate a comprehensive evaluation of the proposed HOCLF-HOCBF-QP controller. Starting from pure path tracking, the controller was shown to achieve stable and accurate path following capability. When static obstacles were introduced, the controller successfully balanced safety and path tracking, both in reference path and reference end point scenarios. Extending to dynamic obstacles, the controller maintained safe distances while performing effective avoidance maneuvers and returning to the desired trajectory. These results reflect the controller's capability to handle emergency conditions in complex driving environments. In the following sections, we further validate its performance by applying it to representative FARS bicyclist crash scenarios.

## *FARS Bicyclist Crash Cases Study*

To further evaluate proposed low-level HOCLF-HOCBF-QP controller's collision avoidance capability, we replicate realistic FARS bicyclist crash scenarios, which capture severe and representative real-world bicyclist accidents. By testing the controller in these scenarios, we can comprehensively evaluate its capability to prevent such accidents in real life.

### **FARS 210: Motorist left turn, bicyclist failed to yield sign**

We first replicate realistic FARS210 bicyclist crash scenarios, which involve a motorist attempting a left turn while a bicyclist simultaneously ignores a yield sign and proceeds forward. Although such events occur only occasionally, they represent critical traffic situations where bicyclists violate traffic regulations. At the same time, motorists making left turns may fail to properly account for approaching bicyclists from the side, substantially increasing the risk of severe collisions.

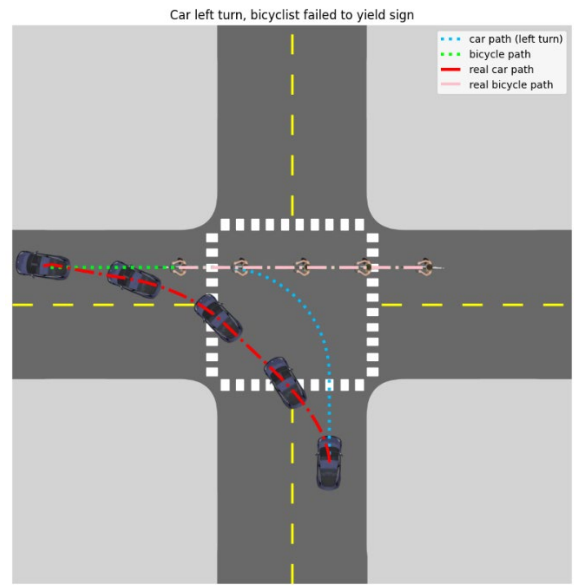

Figure 8. Snapshot of FARS210 Simulation with HOCLF-HOCBF-QP Collision Avoidance Controller

Figure 8 demonstrates the simulation results of the FARS210 scenario under the control of proposed HOCLF-HOCBF-QP controller, which enables ego-vehicle to perform effective path-following and collision avoidance maneuvers. In the figure, the blue dashed line and green dashed line represent the reference paths for the vehicle and the bicyclist, respectively. The red dashed line shows the actual trajectory of the ego-vehicle, while the pink dashed line corresponds to the bicyclist's motion.

From the figure we can observe that if the ego-vehicle were to strictly follow its original reference path, it would inevitably collide with the bicyclist approaching from the side, leading to a severe accident. However, the additional HOCBF safety guarantees constraint actively enforces collision avoidance by forcing the ego-vehicle to deviate from its original trajectory. Meanwhile, the HOCLF path tracking constraint ensures that this deviation remains minimal, maintaining accurate path tracking while guaranteeing safety. The snapshots of the vehicle clearly demonstrate how it smoothly adjusts its left-turn maneuver to avoid the bicyclist and then gradually return toward a safe trajectory without collision.

This example highlights the effectiveness of integrating HOCLF (for accurate path tracking) with HOCBF (for safety guarantees) in the QP framework, showing that the controller can autonomously override potentially unsafe trajectories and ensure safe maneuvering in real-world crash scenarios.

### **FARS 220: Motorist Left Turn / Merge**

Next, we replicate the realistic FARS220 bicyclist crash scenario, in which the ego-vehicle continues to move straight while the bicyclist suddenly changes lanes and attempts to merge in front of the vehicle. Compared with the FARS210 left-turn case, this scenario occurs more frequently in practice, as bicyclists typically lack rear-view mirrors and may forget to check for approaching vehicles before starting a lane change. Such inattentive maneuvers frequently lead to conflicts with following vehicles and can cause severe collisions if the vehicle cannot properly manage to brake.

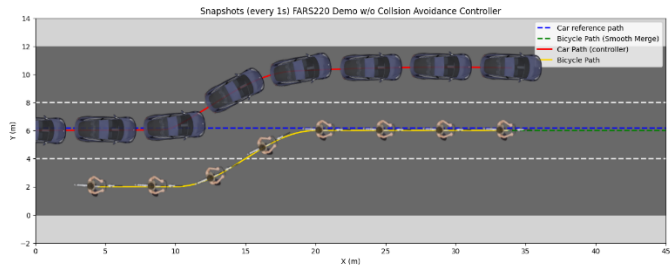

Figure 9. Snapshot of FARS220 Simulation with HOCLF-HOCBF-QP Collision Avoidance Controller

Figure 9 demonstrates the simulation results of the FARS220 scenario under the control of HOCLF-HOCBF-QP low-level controller, which allows the ego-vehicle to perform path-following and collision avoidance simultaneously. In the figure, the blue dashed line and green dashed line represent the reference paths for the vehicle and the bicyclist, respectively. The red solid line shows the actual trajectory of the ego-vehicle, while the yellow solid line corresponds to the bicyclist's motion.

In this case, the bicyclist suddenly changes lanes and cuts in front of the ego-vehicle. If the ego-vehicle were to strictly follow its original reference path or fail to slow down, it would inevitably collide with the bicyclist, leading to a severe crash. Such collisions are particularly dangerous because the impact from behind may cause the bicyclist to suddenly accelerate forward, lose balance, or, in the worst case, be run over by the vehicle. However, the HOCBF safety constraint forces the ego-vehicle to deviate from its original lane and perform a single-lane-change maneuver to avoid collision. Meanwhile, the HOCLF stability constraint ensures that this deviation remains minimal, thereby maintaining path-tracking performance while guaranteeing safety. The snapshots clearly demonstrate how the ego-vehicle dynamically adjusts its trajectory to avoid the bicyclist and subsequently stabilizes onto a safe path.

**FARS 310: Bicyclists Failed to Yield, Midblock**

Finally, we replicate the realistic FARS310 bicyclist crash scenario, in which the ego-vehicle continues straight while a bicyclist suddenly appears from the roadside and enters the traffic lane. Compared with the previous two FARS cases, this scenario is both more common and considerably more challenging to avoid. This type of case can occur in many different settings, such as bicyclists entering from building driveways, parking lots, or sidewalks and often without checking for oncoming vehicles. Such cases require that drivers maintain continuous awareness of sidewalk and roadside conditions. When such an emergency situation unfolds, drivers must perform immediate emergency braking and/or lane changing to prevent a collision. This case offers a comprehensive evaluation of the controller's capability to handle sudden emergencies in a highly complicated and dynamic traffic environment.

Figure 10 demonstrates the simulation results of the FARS310 scenario using the HOCLF-HOCBF-QP low-level controller, which enables both path-following and collision avoidance. In the figure, the blue dashed line and green dashed line represent the reference paths for the vehicle and the bicyclist, respectively. The red dashed line shows the actual trajectory of the ego-vehicle, while the pink dashed line corresponds to the bicyclist's motion.

In this case, the bicyclist suddenly enters the roadway from the sidewalk, cutting directly across the ego-vehicle's path. If the vehicle had continued along its original route or failed to brake in time, a side collision would be unavoidable. Such crashes are especially dangerous, as the impact could cause the bicyclist to overturn and potentially be run over. To mitigate this risk, the HOCBF forces the vehicle to leave its lane and execute an emergency evasive maneuver, while the HOCLF limits the deviation to preserve path-following performance. The snapshots show how the ego-vehicle smoothly adjusts its trajectory to avoid the bicyclist and then stabilizes back onto a safe path.

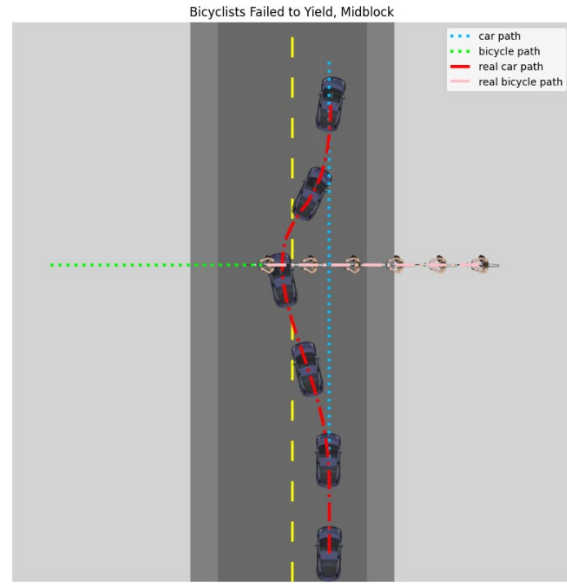

Figure 10. Snapshot of FARS310 Simulation with HOCLF-HOCBF-QP Collision Avoidance Controller

The most appropriate response to this emergency scenario in practice would be to stop, as emergency braking is the most reliable way to handle a bicyclist suddenly entering from the roadside. However, our focus here is on evaluating the low-level controller. In a full autonomous driving framework, the low-level controller typically operates in conjunction with a high-level decision-making module. This means that even if the high-level controller makes a suboptimal or unsafe decision, such as continuing forward despite the bicyclist's intrusion, the additional safety layer introduced by the HOCBF constraint ensures that the low-level controller can still enforce collision avoidance and guarantee safety, which makes this method suitable for use in an autonomous driving framework.

## Conclusion

Bicyclists are highly vulnerable road users due to their small size, lower visibility, and their frequent presence in vehicle blind spots. Therefore, ensuring bicyclist safety remains a fundamental challenge in autonomous driving system development. To address this challenge, we proposed an HOCLF-HOCBF-QP framework. Designed as a low-level controller, it can be integrated with high-level planning modules to provide both trajectory tracking and safety guarantees. Unlike traditional path-tracking approaches that primarily enforce trajectory following, the HOCLF-HOCBF-QP controller incorporates an additional safety constraint through the HOCBF, ensuring the simultaneous achievement of path tracking and safety.

The controller's effectiveness was validated through a series of progressively complex evaluation experiments. Starting with basic path-tracking tasks, we demonstrated stable path following capability. Extending to static and dynamic obstacle scenarios, the controller successfully balanced reference tracking and safety. Finally, by replicating and testing our controller in three representative FARS bicyclist crash scenarios, we showed that the controller can autonomously track the desired path and perform collision avoidance maneuvers in a dynamic traffic environment to avoid collisions with bicyclists during emergency situations.

Together, these results confirm that the HOCLF-HOCBF-QP controller provides a reliable low-level safety mechanism that complements high-level decision-making, significantly enhancing the safety of bicyclists and other vulnerable road users in complex traffic environments.

## Future Work

Despite the advancements presented in this study, the proposed HOCLF-HOCBF-QP low-level controller still has several limitations that need further investigation and adjustment.

First, as an optimization-based approach, the controller may demand substantial computational resources when deployed in highly complex environments. This raises concerns about its real-time capability in autonomous driving systems. To address this challenge, future work will incorporate Hardware-in-the-Loop (HIL) and Vehicle-in-Virtual-Environment (VVE) testing methods [46] to comprehensively evaluate the controller's real-time performance and ensure its feasibility for real-world implementation.

Second, the effectiveness of the controller depends heavily on the choice of hyperparameters. In this study, hyperparameters were selected based on empirical knowledge and simulation study, which proved sufficient for the presented scenarios. However, identifying the most robust hyperparameter set across varying conditions remains an open problem. Developing algorithms that capable of dynamically tuning hyperparameters in response to real-world conditions will be an important direction for future research.

Although the incorporation of HOCBF provides an additional safety guarantee at the low level, potential conflicts may exist when combing with the supervised learning or deep reinforcement learning based high-level planner. For example, situations may occur where the high-level decision-making agent sends a command that contradicts the safety-enforced low-level response (e.g., the high-level directing a left maneuver while the low-level enforces a right adjustment). Future work must investigate methods to integrate high-level decision making with low-level safety constraints to avoid such conflicts.

Future work can also focus on use of V2X communication for enhanced VRU perception [47], using parameter space based robust low level control [48] and consider the effect of time delays on performance [49] and use delay tolerant low level controls [50].

Syst., vol. 23, no. 6, pp. 4909–4926, June 2022, doi: 10.1109/TITS.2021.3054625.

[14] F. Ye, S. Zhang, P. Wang, and C.-Y. Chan, "A Survey of Deep Reinforcement Learning Algorithms for Motion Planning and Control of Autonomous Vehicles," in *2021 IEEE Intelligent Vehicles Symposium (IV)*, July 2021, pp. 1073–1080. doi: 10.1109/IV48863.2021.9575880.

[15] Z. Zhu and H. Zhao, "A Survey of Deep RL and IL for Autonomous Driving Policy Learning," *IEEE Trans. Intell. Transp. Syst.*, vol. 23, no. 9, pp. 14043–14065, Sept. 2022, doi: 10.1109/TITS.2021.3134702.

[16] Z. Zhang, C. Peng, M. Zhu, E. Yurtsever, and K. A. Redmill, "An Uncertainty-Weighted Decision Transformer for Navigation in Dense, Complex Driving Scenarios." Accessed: Sept. 30, 2025. [Online]. Available: https://arxiv.org/abs/2509.13132

[17] Z. Cao et al., "Reinforcement Learning based Control of Imitative Policies for Near-Accident Driving," June 30, 2020, *arXiv*: arXiv:2007.00178. doi: 10.48550/arXiv.2007.00178.

[18] A. Aksjonov and V. Kyrki, "A Safety-Critical Decision-Making and Control Framework Combining Machine-Learning-Based and Rule-Based Algorithms," *SAE Int. J. Veh. Dyn. Stab. NVH*, vol. 7, no. 3, pp. 10-07-03–0018, June 2023, doi: 10.4271/10-07-03-0018.

[19] "Deep reinforcement-learning-based driving policy for autonomous road vehicles - Makantasis - 2020 - IET Intelligent Transport Systems - Wiley Online Library." Accessed: Oct. 24, 2023. [Online]. Available: https://ietresearch.onlinelibrary.wiley.com/doi/full/10.1049/iet-its.2019.0249

[20] F. Merola, F. Falchi, C. Gennaro, and M. Di Benedetto, "Reinforced Damage Minimization in Critical Events for Self-driving Vehicles:," in *Proceedings of the 17th International Joint Conference on Computer Vision, Imaging and Computer Graphics Theory and Applications*, Online Streaming, --- Select a Country ---: SCITEPRESS - Science and Technology Publications, 2022, pp. 258–266. doi: 10.5220/0010908000003124.

[21] H. Chen, F. Zhang, and B. Aksun-Guvenc, "Collision Avoidance in Autonomous Vehicles Using the Control Lyapunov Function–Control Barrier Function–Quadratic Programming Approach with Deep Reinforcement Learning Decision-Making," *Electronics*, vol. 14, no. 3, p. 557, Jan. 2025, doi: 10.3390/electronics14030557.

[22] H. Chen and B. Aksun-Guvenc, "Hierarchical Deep Reinforcement Learning-Based Path Planning with Underlying High-Order Control Lyapunov Function—Control Barrier Function—Quadratic Programming Collision Avoidance Path Tracking Control of Lane-Changing Maneuvers for Autonomous Vehicles," *Electronics*, vol. 14, no. 14, p. 2776, Jan. 2025, doi: 10.3390/electronics14142776.

[23] Y. Ding, H. Zhong, Y. Qian, L. Wang, and Y. Xie, "Lane-Change Collision Avoidance Control for Automated Vehicles with Control Barrier Functions," *Int. J. Automot. Technol.*, vol. 24, no. 3, pp. 739–748, June 2023, doi: 10.1007/s12239-023-0061-2.

[24] I. Jang and H. J. Kim, "Safe Control for Navigation in Cluttered Space Using Multiple Lyapunov-Based Control Barrier Functions," *IEEE Robot. Autom. Lett.*, vol. 9, no. 3, pp. 2056–2063, Mar. 2024, doi: 10.1109/LRA.2024.3349917.

[25] A. A. Nasab and M. H. Asemani, "Control of Mobile Robots Using Control Barrier Functions in Presence of Fault," in *2023 9th International Conference on Control, Instrumentation and Automation (ICCIA)*, Dec. 2023, pp. 1–6. doi: 10.1109/ICCIA61416.2023.10506347.

[26] A. Alan, A. J. Taylor, C. R. He, A. D. Ames, and G. Orosz, "Control Barrier Functions and Input-to-State Safety With Application to Automated Vehicles," *IEEE Trans. Control Syst. Technol.*, vol. 31, no. 6, pp. 2744–2759, Nov. 2023, doi: 10.1109/TCST.2023.3286090.

[27] S. Nageshrao, E. Tseng, and D. Filev, "Autonomous Highway Driving using Deep Reinforcement Learning," Mar. 29, 2019, *arXiv*: arXiv:1904.00035. doi: 10.48550/arXiv.1904.00035.

[28] Z. Zhang, C. Peng, E. Yurtsever, and K. A. Redmill, "Bootstrapping Reinforcement Learning with Sub-optimal Policies for Autonomous Driving," Sept. 04, 2025, *arXiv*: arXiv:2509.04712. doi: 10.48550/arXiv.2509.04712.

[29] "Evaluation of a Simple Pure Pursuit Path-Following Algorithm for an Autonomous, Articulated-Steer Vehicle." Accessed: Oct. 23, 2023. [Online]. Available: https://doi.org/10.13031/aea.30.10347

[30] Y. Chen and J. J. Zhu, "Pure Pursuit Guidance for Car-Like Ground Vehicle Trajectory Tracking," presented at the ASME 2017 Dynamic Systems and Control Conference, American Society of Mechanical Engineers Digital Collection, Nov. 2017. doi: 10.1115/DSCC2017-5376.

[31] G. M. Hoffmann, C. J. Tomlin, M. Montemerlo, and S. Thrun, "Autonomous Automobile Trajectory Tracking for Off-Road Driving: Controller Design, Experimental Validation and Racing," in *2007 American Control Conference*, July 2007, pp. 2296–2301. doi: 10.1109/ACC.2007.4282788.

[32] B. A. Güvenç, L. Güvenç, and S. Karaman, "Robust MIMO disturbance observer analysis and design with application to active car steering," *Int. J. Robust Nonlinear Control*, vol. 20, no. 8, pp. 873–891, 2010, doi: 10.1002/rnc.1476.

[33] L. Güvenç, B. Aksun-Güvenç, B. Demirel, and M. T. Emirler, *Control of Mechatronic Systems*. IET Digital Library, 2017. doi: 10.1049/PBCE104E.

[34] H. Wang, S. Y. Gelbal and L. Guvenc, "Multi-Objective Digital PID Controller Design in Parameter Space and its Application to Automated Path Following," in *IEEE Access*, vol. 9, pp. 46874-46885, 2021, doi: 10.1109/ACCESS.2021.3066925.

[35] X. Chen, X. Liu, and M. Zhang, "Autonomous Vehicle Lane-Change Control Based on Model Predictive Control with Control Barrier Function," in *2024 IEEE 13th Data Driven Control and Learning Systems Conference (DDCLS)*, May 2024, pp. 1267–1272. doi: 10.1109/DDCLS61622.2024.10606562.

[36] A. Thirugnanam, J. Zeng, and K. Sreenath, "Safety-Critical Control and Planning for Obstacle Avoidance between Polytopes
Page 9 of 10

## Acknowledgments

This work was funded in part by Carnegie Mellon University's Safety21 National University Transportation Center, which is sponsored by the US Department of Transportation.

The authors thank NVIDIA for its GPU donations.

The authors thank the Automated Driving Lab in Ohio State University for supporting this work.